\documentclass[10pt]{iopart}

\begin{document}

\title{$K^+/\pi^+$ Probes of Jet Quenching in AA Collisions}

\author{
P. L\'evai$^{\dag \ddag}$, G. Papp$^{\S \P}$, G. Fai$^{\P}$, 
and M. Gyulassy$^{\ddag}$} 
\address{
\dag\  KFKI Research Institute for Particle and Nuclear Physics, \\
       PO Box 49, Budapest, 1525, Hungary\\
\ddag\ Department of Physics, Columbia University, \\
       538 West 120th Street, New York, NY 10027, USA\\
\S \   HAS Research Group for Theoretical Physics, \\
      E\"otv\"os University, P\'azm\'any P. 1/A, Budapest 1117, Hungary \\
$\P$ Physics Department, Kent State University, \\
   Kent OH 44242, USA 
}

\begin{abstract}
Non-abelian energy loss in quark gluon plasma is shown to lead to novel 
hadron ratio suppression patterns in ultrarelativistic nuclear collisions.
We apply  recent (GLV) estimates for the gluon radiative energy loss,
which increases linearly with  the jet energy up to $E<20$ GeV and depends
 quadratically on  the nuclear radius, $R$.
The $K^+/\pi^+$ ratio is found to be  most sensitive to
 the initial density of the plasma. 
\end{abstract}

Energy loss of high energy quark and gluon jets penetrating  dense
matter produced in ultrarelativistic heavy ion collisions leads to jet 
quenching and thus  probes the quark-gluon plasma formed in those 
reactions~\cite{gptw,mgxw92}.
The non-abelian radiative energy loss, $\Delta E(E,L)$, suppresses the moderate
$2-3 \ {\rm GeV} < p_T < 10-15 \ {\rm GeV}$ distributions of hadrons in a way 
that can also influence the jet fragmentation pattern into different
flavor hadrons. This is because quark and gluon jets suffer different
energy losses proportional to their color Casimir factors $(4/3,3)$. 
First estimates~\cite{gptw,mgxw92} suggested that 
$\Delta E \approx 1-2 \ {\rm GeV}(L/{\rm fm})$ would depend linearly
on the plasma thickness, $L$, as in abelian electrodynamics.
In BDMS~\cite{bdms8} non-abelian (radiated gluon final state
interaction) effects were shown to lead to  a quadratic
dependence on $L$ with a much larger magnitude of $\Delta E$.
The analysis of jet-quenching in $Pb+Pb$ at $\sqrt{s}=20$ AGeV, however, 
indicated a negligible energy loss at that energy~\cite{XNWA98,MGPL}.
In GLV~\cite{glv2,glv2b} finite kinematic constraints were found to reduce 
 greatly the energy loss at moderate jet energies and we obtained
$\Delta E(E,L)\sim E \cdot (L/6\;{\rm fm})^2$, 
Here we apply the GLV energy loss to estimate more 
quantitatively the effect of jet quenching on the hadronic ratios at
moderate $p_\perp<10$ GeV at the Relativistic Heavy Ion Collider (RHIC)
energy, $\sqrt{s} = 130 \  {\rm AGeV}$.
We illustrate the jet quenching effects for a generic plasma with 
an average screening scale $\mu=0.5$~GeV, 
strong coupling $\alpha_s=0.3$, and an average 
gluon mean free path  $\lambda_g=1$~fm.  We introduce the opacity as the 
mean number of jet scatterings, $\bar{n}=L/\lambda$, which is assumed to 
be a finite number.

An advantage of looking into particle ratios is that uncertainties in the
absolute normalization due to acceptance tend to cancel.
The disadvantage is of course that high $p_\perp$ particle identification
is increasingly difficult.
As we have shown in Ref.~\cite{ktopi}, the kinematically
suppressed GLV energy loss turns out to depend
 approximately linearly on the jet energy, $E$. 
This linear dependence, $\Delta E\propto E$, leads
to only a  very weakly $p_\perp$ dependent
suppression of the transverse momentum distributions for
the kinematic range accessible experimentally at RHIC.
We focus on the $p_\perp$ dependence of the particle ratios 
in order to help  pin down  the jet quenching mechanism.
Our primary candidate is the  measurable $K^+/\pi^+$ ratio. 

In order to investigate the influence of the GLV  energy-dependent radiative 
energy loss on  hadron production, we apply a perturbative QCD (pQCD) based 
description of $Au+Au$ collisions, including energy loss prior to hadronization.
First, we check  that the applied pQCD description reproduces data on pion and 
kaon production in $p+p$ collision.  Our results are based on a leading order 
(LO) pQCD analysis.  Detailed discussion of the formalism is published 
elsewhere \cite{PLF00,YFP02}. 

Our pQCD calculations incorporate the parton transverse momentum 
(``intrinsic $k_T$'') via a Gaussian transverse momentum distribution 
$ g(\vec{k}_{T})$ (characterized by
the width $\langle k_T^2 \rangle $) \cite{PLF00,YFP02,XNWint} 
with the usual convolution of the parton distribution functions (PDF)
$f_{a/p}$, partonic cross sections and fragmentation functions
(FF)  $D_{h/c}$, as: 
\begin{eqnarray}
\label{fullpipp}
&&E_{h}\frac{d\sigma_h^{pp}}{d^3p} =\!
        \sum_{abcd}\!  
        \int\!\!dx_1 dx_2 dz_c d^2k_{T,a}d^2k_{T,b}\
        g(\vec{k}_{T,a}) g(\vec{k}_{T,b}) \times
        \nonumber \\
        && \ \ \ f_{a/p}(x_1,Q^2) f_{b/p}(x_2,Q^2)\
             \frac{d\sigma}{d{\hat t}}
   \frac{D_{h/c}(z_c,{\widehat Q}^2)}{\pi z_c^2}{\hat s}
\delta({\hat s}+{\hat t}+{\hat u})  \, .
\end{eqnarray}

Here we use LO PDF from the MRST98 parameterization \cite{MRS98} and a
LO FF~\cite{BKK95}.
The applied scales are $Q=p_{\bf c}/2$ and ${\widehat Q}= p_T/2z_c$.

Utilizing available high transverse-momentum ($2 < p_T < 10$ GeV)
$p+p$ data on pion and kaon production at $19 < \sqrt{s} < 63$ GeV 
we can determine the best fitting energy-dependent 
$\langle k_T^2 \rangle$ parameter 
(see Ref.~\cite{PLF00,YFP02} for further details on $\langle k_T^2 \rangle$).
One can conclude about the $K^+/\pi^+$ ratio that, while agreement is 
reasonable for $p_T \geq 4$ GeV, there is a systematic  discrepancy at 
smaller transverse momenta.
High statistics $p+p$ data at RHIC will be essential to establish an accurate 
baseline to which $A+A$ must be compared.
Now we turn to the calculation for $Au+Au$ collision at RHIC energy,
$\sqrt{s} =130$ AGeV.  As a first approximation, let us consider slab geometry, 
neglecting radial dependence.  We include the isospin asymmetry and the nuclear
modification (shadowing) into the nuclear PDF \cite{HIJ}.

The value of the $\langle k_T^2 \rangle$ of the transverse component of the PDF
will be increased by multiscattering effects in $A+A$ collisions. Two limiting 
cases were investigated with (i) a large number of rescatterings \cite{XNWint}
and (ii) a small number of rescatterings (``saturated Cronin effect'') 
~\cite{PLF00,YFP02}.
Here we consider the influence of jet-quenching on the hadron spectra,
leaving the inclusion of the multiscattering effect for future work.
In this simplified calculation, jet quenching reduces  the energy
of the jet before fragmentation. We concentrate on $y_{cm}=0$,
 where the jet transverse
momentum before fragmentation is shifted by the energy loss,
$p_c^*(L/\lambda) = p_c - \Delta E(E,L)$. This 
shifts the $z_c$ parameter in the integrand
to $z_c^* = z_c /(1-\Delta E/p_c)$.
The applied scale in the FF
is  similarly modified,
${\widehat Q} = p_T/2z_c^*$, while
for the elementary hard reaction the scale
remains $Q=p_c/2$.

With these approximations the invariant cross section of hadron
production in central $A+A$ collision is given
(in a somewhat abbreviated notation) by
\begin{eqnarray}
\label{fullaa}
E_{h}\frac{d\sigma_h^{AA}}{d^3p} &=& B 
        \sum_{abcd}\!  
        \int\!\!dx_1 dx_2 dz_c d^2k_{T,a}d^2k_{T,b}\
        g(\vec{k}_{T,a}) g(\vec{k}_{T,b}) \times
        \nonumber \\
        && \ \ \ \   f_{a/A} \ f_{b/A} \ 
             \frac{d\sigma}{d{\hat t}} \frac{z^*_c}{z_c}
   \frac{D_{h/c}(z^*_c,{\widehat Q}^2)}{\pi z_c^2} {\hat s}
\delta({\hat s}+{\hat t}+{\hat u})  \, . 
\end{eqnarray}
where $B  = 2 \pi \int_0^{b_{max}} b~db~T_{AA}(b)$ 
with $b_{max}= 4.7$ fm for the 10 \% most central $Au+Au$ collision
and $T_{AA}(b)$ is the thickness function.
The factor $z^*_c/z_c$ appears because of  the in-medium 
modification of the fragmentation function \cite{WaHu97}.
Thus, the invariant cross section 
(\ref{fullaa}) will depend on the average
opacity or collision number, ${\bar n} = L/\lambda_g$.

We note that the GLV energy-dependent energy loss leads to a rather 
structureless downward shift of the single inclusive yields of all hadrons 
because $\Delta E/E$ is approximately constant. This is in contrast to the 
results in \cite{XNWint}, where an energy dependent fractional energy loss,
$\Delta E/E=(0.5 \  {\rm GeV}/E)(L/{\rm fm})$, was used.

More generally, the quenching factor depends also on fluctuations of 
energy loss that occurs because a single collision dominates the energy loss. 
We are presently testing the effect of fluctuations using the normalized
$f(\delta E)\propto \theta(\delta E- \delta E_0)/\delta E^2$ 
distribution of energy loss from~\cite{glv2} 
and $\delta E_0$ fixed by $\langle \delta E\rangle =\Delta E_{GLV}(E,L)$. 
Preliminary results indicate, that fluctuations simply renormalize the 
effective opacity that is needed to obtain a fixed quenching factor.
The reason that this simplification occurs in spite of the large effect of
fluctuations is the approximate energy independence of the GLV {\em fractional}
energy loss in the relevant jet energy range.  Since other effects, such as 
expansion, also renormalize the value of the effective static opacity, 
we continue here to illustrate the effects of 
jet quenching simply varying $L/\lambda_g$ as discussed above. 

To understand the origin of the different dependence of the $\pi$ and
$K$ quenching factors on $L/\lambda_g$, we consider the gluon contribution
to the hadron yield for charged pions and kaons. For simplicity, we illustrate
the gluon contribution to the non-quenched ratios only. Fig. 1.
displays (top panel) the relative contribution of gluon fragmentation
to the production of pions and kaons as a function of $p_T$. It can be seen 
that gluons dominate pion production while (valence) quarks dominate $K^+$ 
production, especially at lower $p_T$ values. 
Due to the dependence of jet energy loss on the color Casimir factor,
pions will experience stronger suppression than kaons for
the same value of ${\bar n} = L/\lambda$. There is 
a $p_T$ dependence of this effect because the quark and 
gluon fragmentation functions differ. While $\pi^+$ and $\pi^-$
show an almost identical relation between quark and gluon
fragmentation contribution, there is a  difference between 
$K^+$ and $K^-$ in accordance with the different production
mechanism of $K^+$ and $K^-$ \cite{LevPet}.
Our main result is the prediction
that there could be a factor of 2-3 enhancement in the $K^+/\pi^+$ ratio
in the $p_T\sim 3-6$ GeV range for effective opacity ${\bar n}=4$. 
The recently reported suppression of the $\pi^0$ spectra 
in Ref. \cite{QM01David} corresponds to an effective static opacity 
in the range ${\bar n} \approx 3-4$ \cite{QM01LP}.
 
In summary, we applied the GLV parton energy loss~\cite{glv2,glv2b} in LO pQCD
to pion and kaon production in nuclear collisions. The energy dependence
of $\Delta E_{GLV}$ leads to a weak $p_T$ dependent  but
 {\em flavor dependent}
quenching factor that increases with the effective opacity of the plasma.
The enhancement of the
$K^+/\pi^+$  ratio in the $2< p_\perp<10 $ GeV range 
was found to be the most sensitive to the gluon opacity at RHIC energies.

\ack
This work was supported by the U.S. DOE 
under DE-FG02-86ER40251, DO-AC03-76SF00098 and 
DE-FG02-93ER40764, by the U.S. NSF under INT-0000211,
by the BCPL in Bergen and OTKA No. T029158, T032796. 

\newpage

\begin{center}
\vspace*{10.0cm}
\includegraphics{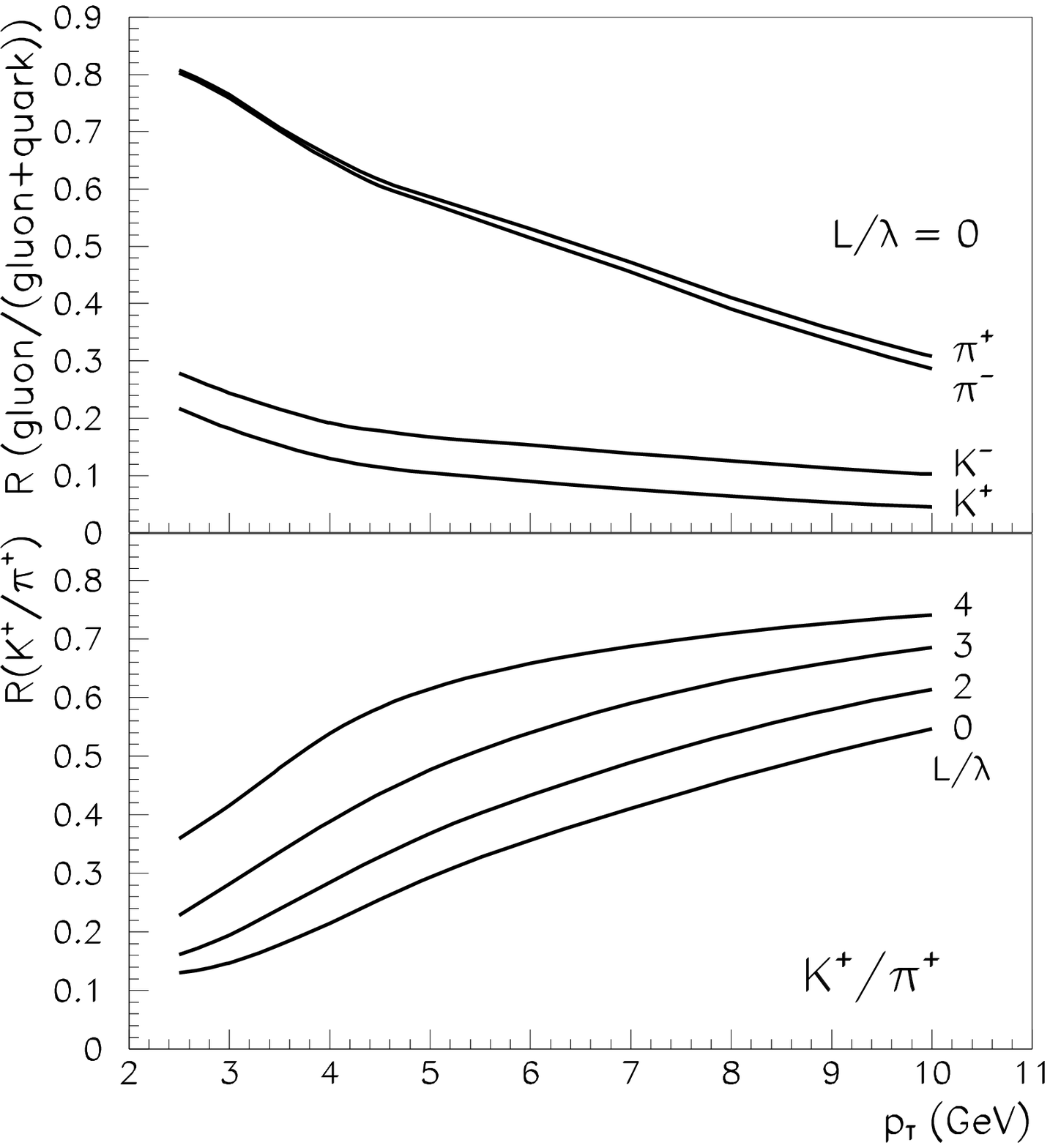}
\vspace{-1.0cm}
\end{center}
\begin{center}
\begin{minipage}[t]{13cm}  
      { FIG. 1.} {\small Top panel: 
Relative contribution of gluon fragmentation 
to charged pion and kaon production at $\sqrt{s}=130$ AGeV
without jet quenching. 
Bottom panel:
$K^+/\pi^+$ ratio in $ Au+Au \rightarrow h+ X$
collision at $\sqrt{s}=130 \ AGeV$
without jet-quenching $({\bar n}=0)$ 
and with it,  ${\bar n}=2,3,4$.
} 
\end{minipage}
\end{center}

\vfill\eject

\end{document}